\documentclass[aps,prb,reprint,amsmath,amssymb,groupedaddress]{revtex4-2}

\usepackage{graphicx}
\usepackage{dcolumn}
\usepackage{bm}
\usepackage{color}
\usepackage{siunitx}
\usepackage{todonotes}
\usepackage{threeparttable}
\usepackage{gensymb}
\usepackage{ulem}

\definecolor{imcolor}{rgb}{0.5,0.,0.5}				

\definecolor{ohcolor}{rgb}{0.,0.5,0.5}

\definecolor{hpcolor}{rgb}{0.5,0.5,0}

\begin{document}

\title{Theoretical design of the large topological magnetoelectric effect in the Co-intercalated NbS$_2$ structure}

\author{Hyowon Park$^{1,2}$ and Ivar Martin$^{2}$}

\affiliation{$^1$Department of Physics, University of Illinois at Chicago, Chicago, IL 60607, USA,\\
$^2$Materials Science Division, Argonne National Laboratory, Argonne, IL, 60439, USA
 }

\date{\today}

\begin{abstract}
A triangular Co-ion lattice intercalated between 1-H NbS$_2$ layers can exhibit a large anomalous Hall effect (AHE) due to the finite scalar spin chirality originating from the non-coplanar $3q$ ordering of Co spins.
This large AHE occurs when the scalar spin chirality is uniform in all Co layers, as indeed found in the Co$_{1/3}$NbS$_2$ case [Phys. Rev. Mater. {\bf6}, 024201 (2022)].
However, if the spin chirality were staggered with the opposite signs in the adjacent Co layers, the net AHE would disappear, yielding instead the topological magneto-electric effect. Here, we theoretically verify that a transverse electric field generates a finite orbital magnetization under such conditions, consistent with the axion-like coupling. Using first-principles calculations, we show that the resulting magneto-electric coupling, $\alpha^{zz}$ can be as large as 0.9 $e^2/2h$. We also demonstrate that the inter-layer magnetic coupling in these materials can be tuned by strain, enabling the switching between the AHE and the axionic states.  
\end{abstract}

\maketitle

\section{Introduction}

The topological magneto-electric effect has been  proposed in $3D$ topological insulator (TI) ~\cite{PhysRevB.92.085113} and ferromagnetic insulator (FI)-TI-FI heterostructures~\cite{PhysRevB.92.081107}. The basic physics of the effect can be understood in terms of the surface anomalous Hall current induced by an applied electric ($\mathbf{E}$) field, which produces an orbital magnetization. This electric-field induced magnetization, or the complementary magnetic field ($\mathbf{B}$) induced electric polarization can be interpreted as originating from the effective action that includes the axion coupling term~\cite{PhysRevB.78.195424,PhysRevLett.102.146805,Nenno2020}:
\begin{equation}
\label{eq:dmorb}
S_\theta=\frac{\alpha}{4\pi^2}\int d^3x \: dt\: \theta \: \mathbf{E}\cdot\mathbf{B},
\end{equation}
where $\alpha (=e^2/\hbar c)$ is the fine-structure constant and $\theta$ is the axion angle.


For systems with either time-reversal or inversion symmetries,  $\theta$ is constrained to be 0 or $\pi$.
When both of these symmetries are broken, $\theta$ is no longer quantized, becoming a degree of freedom tunable  magnetically, electrically, or optically. However, the value of $\theta$ in these cases tends to be small~\cite{10.1063/1.1728630}. 
A proposal by Li $et$ $al.$~\cite{Li2010} hypothesized a Fe doped Bi$_2$Se$_3$ as a system where a sizable non-quanitized $\theta$ could appear. However, no successful synthesis of this material has been reported.

Recently, it was realized that thin films of MnBi$_2$Te$_4$ can also implement non-quantized axion coupling.  Several dc and optical frequency measurements  have indeed confirmed electromagnetic coupling to and control of the axion field~\cite{Qiu2023,Qiu2025}.
MnBi$_2$Te$_4$ has a layered structure along with the collinear antiferromagnetic ordering of Mn bi-layers~\cite{Otrokov2019}. 
The first-principles calculation has shown that the axion coupling in this material is dominated by the orbital contribution which is directly related to the topological band structure and the axion coupling term can be estimated from the layer-dependent Berry curvature calculation~\cite{Qiu2025}.

Another example of an effect rooted in band topology is anomalous Hall effect (AHE)~\cite{nagaosa_anomalous_2010}. 
It can originate from the interplay of magnetic order, crystal structure, and spin-orbit coupling. 
One route involves the non-coplanar magnetic order with uniform scalar spin chirality~\cite{martin_scalar_2008}, which can induce band Berry curvature, leading to finite Hall effect even in the absence of spin-orbit interaction and with a negligible uniform magnetization. 
Experimental evidence indicates that this mechanism is responsible for the very large AHE in Co-intercalated NbS$_2$~\cite{Gu2025} and TaS$_2$~\cite{Park2023}.  The Berry curvature can be computed using first-principles and the obtained anomalous Hall conductivity can be comparable to the experimentally measured value~\cite{PhysRevMaterials.6.024201}.

In this paper, we point out that a closely related magnetic state with 
{\it staggered} layer scalar spin chirality will have a strong magneto-electric coupling. As we have found previously~\cite{PhysRevMaterials.6.024201}, this state is energetically competitive with other magnetic orders in several intercalated transition-metal dichalcogenides. Here we also show that the inter-layer magnetic coupling, which controls whether the uniform or staggered spin chirality has a lower energy, can be tuned by applying in-plane strain.
To demonstrate the magneto-electric coupling, we simulated the finite electric field ($E_z$) effect by shifting the on-site orbital energies, i.e. $\delta E \sim E_z\cdot d$, where $d$ is the position of the orbital along the $z-$direction.
The large topological magnetoelectric coupling $\alpha^{zz}$ is expected for the anti-chiral (AC) magnetic structure, where the Berry curvature contributions can have the opposite sign between top and bottom layers and the coupling of the electric field to the layer-dependent topological band structures can be optimal.
We calculate this topological magnetoelectric coupling directly from the linear slope of the orbital magnetization vs the electric field based on first-principles and show that the resulting coupling value can be remarkably large ($\sim 0.9\:e^2/2h)$. 
The effect is dominated by the layer-dependent Berry curvature.

\section{Computational Method}
\label{sec:method}

To compute the energetics and band structures, we adopt Vienna Ab-initio Simulation Package (VASP)~\cite{vasp1,vasp2} implemented using the projector augmented wave method~\cite{PAW} and perform density functional theory (DFT)+U calculations. 
We use the Hubbard $U$ value of 5eV and the Hund's coupling $J$ value of 0.8eV for Co ions. This Hubbard $U$ value is estimated from previous linear response calculations~\cite{PhysRevB.105.155114,PhysRevB.107.235149} and has been reported to produce the band structure consistent with the experimental ARPES data~\cite{PhysRevB.109.085110}.
The Perdew-Burke-Ernzerhof (PBE) functional is used for the exchange-correlation functional~\cite{PBE_functional}.
We use the energy cut-off for the plane-wave basis as 400\:eV and a $8\times8\times1$ $k-$point grid for the slab structure.
To compute the magnetic states of collinear and non-collinear AFM, we adopt the $2\times2\times1$ magnetic cell extended from the primitive cell. 
To construct the tight-binding Hamiltonian, we adopt the maximally localized Wannier function method~\cite{Wannier90_2012} using Co $3d$ and Nb $4d_{z^2}$ Wannier orbitals.
The obtained orbital spreads ($\sqrt{\langle r^2\rangle-\langle r\rangle^2}$) for the maximally localized Wannier functions are 3.37\AA\:(Co $d_{z^2}$), 2.79\AA\:(Co $d_{xz}$/$d_{yz}$), 1.89\AA\:(Co $d_{x^2-y^2}$/$d_{xy}$), and 3.07\AA\:(Nb $d_{z^2}$). These Wannier orbitals can properly represent the layer-dependent physical quantities since they are centered at each atomic site and well localized.
The layer-dependent Berry curvature $\Omega_n^{T(B)}(\mathbf{k})$ can be obtained by applying the square of the overlap between the Kohn-Sham orbital $\psi_n(\mathbf{k})$ and the Wannier orbital $\phi^{T(B)}(\mathbf{k})$ centered at the top (T) or the bottom (B) layer: $\Omega_n^{T(B)}(\mathbf{k})\sim \Omega_n(\mathbf{k})\cdot|\langle \psi_n(\mathbf{k})|\phi^{T(B)}(\mathbf{k})\rangle|^2$.
For the Berry curvature and orbital magnetization calculations based on the tight-binding Hamiltonian, we adopt the Wannier-Berry package~\cite{Wannier_berry} using the recursive adaptive refinement method for the $k-$mesh integration.
For the smooth convergence, we used the $500\times500\times1$ $k-$mesh and the temperature smoothing of the Fermi functions at $T=10K$. 
    
\section{The crystal structure}

\begin{figure}[!ht]
\vspace{-0.1cm}
\includegraphics[width=0.7\linewidth]{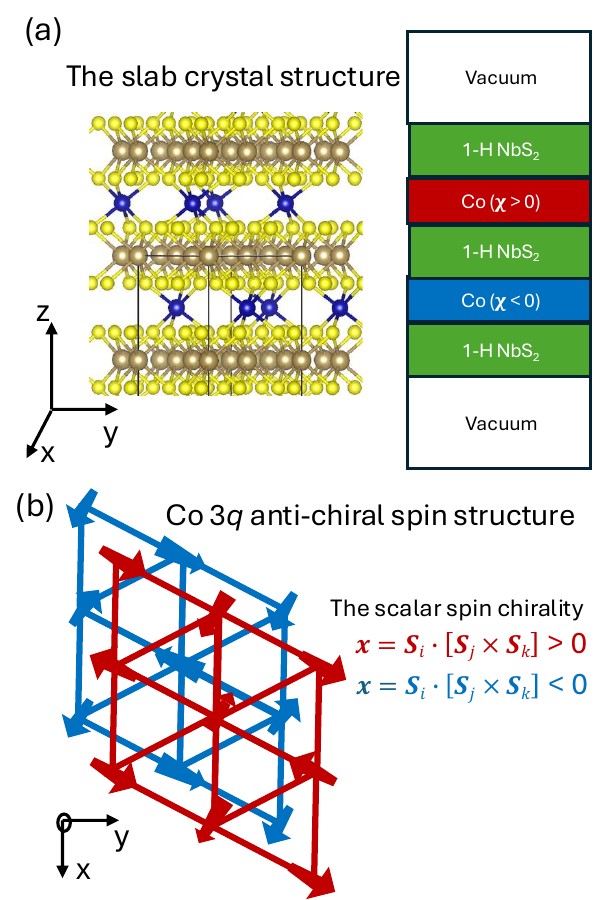}
\caption{ (a) The crystal structure of vacuum/(NbS$_2$)$_3$/Co/  (NbS$_2$)$_3$/Co/(NbS$_2$)$_3$/vacuum thin film using the slab geometry, (b) The schematic spin structure of Co ions in the $3q$ anti-chiral state.  
}
\label{fig:struct}
\end{figure}

Experimentally, Co ions intercalated between the NbS$_2$ layers with the triangular structure (Co$_{1/3}$NbS$_2$) form an antiferromagnetic (AFM) spin structure; yet the material exhibits a very large AHE~\cite{ghimire_large_2018,tenasini_giant_2020}. 
The usual collinear AFM structure typically does not lead to an AHE since the AHE must vanish if the material has a combined symmetry of the time-reversal and the crystal translation.
Instead, it has been argued that the origin of the observed large AHE in Co$_{1/3}$NbS$_2$ is a non-coplanar $3q$ magnetic state; it was indeed subsequently confirmed in the polarized neutron scattering experiments~\cite{Takagi2023}.
Our previous first-principles based calculations confirmed that Co$_{1/3}$NbS$_2$ with the ``tetrahedral" non-coplanar $3q$ magnetic order can generate a large anomalous Hall conductivity comparable to $e^2/h$ per  crystalline layer and each layer can host the finite scalar spin chirality $\chi$ within the triangular plaquette due to the non-coplanar spin structure~\cite{PhysRevMaterials.6.024201}. 
While the Co$_{1/3}$NbS$_2$ compound shows the same  magnetic structure in both Co layers, it is possible that the sign of the scalar spin chirality can be made to alternate from one Co layer to another by tuning the sign of the interlayer spin exchange.
Our previous calculation shows that the total AHE vanishes for this anti-chiral (AC) spin structure since the combined symmetry of the time-reversal and the lattice translation is restored.

To study the transition between the chiral and anti-chiral magnetic states, we design a novel thin-film structure consisting of two  Co layers and three  (top/middle/bottom) NbS$_2$ layers (see Fig.\:\ref{fig:struct}). Similarly to the Co$_{1/3}$NbS$_2$, Co ions form a triangular lattices  intercalated between NbS$_2$ layers. We impose a slab structure along the $z-$direction that includes the top and bottom vacuum layers and also the same 1-H NbS$_2$ termination layers to keep the same chemical environment for both Co layers and similar crystal symmetry as in the bulk Co$_{1/3}$NbS$_2$.
The sulfur ions in the 1-H NbS$_2$ layer break the inversion symmetry, and the entire crystal is non-centrosymmetic, same as the Co$_{1/3}$NbS$_2$ parent material. 
We note that the electronic structure of this thin film structure can be rather different from that of the bulk Co$_{1/3}$NbS$_2$ since it contains three NbS$_2$ layers in the unit cell.
We obtain the lattice constants and internal atomic positions of this thin film by fully relaxing the crystal structure using DFT+U.
The resulting structural parameters are given in Table\:\ref{tbl:energy}.

\section{Strain effect}

\begin{table*}[ht]
    \centering
    \begin{tabular}{|c|c|c|c|c|c|c|c|c|c|c|c|c|} 
    \hline
    Co$_{1/3}$NbS$_2$ & $l_a$ [\AA] & $l_c$ [\AA] & $d$ [\AA] & A-AFM[eV] & FM[eV] & $1q$-AFM1 & $1q$-AFM2 & $2q$-AFM1 & $2q$-AFM2 & $3q$-Chiral & $3q$-Anti\:Chiral \\ [1.0ex] 
    \hline
     Bulk & 5.80 & 11.77 & 5.89 & -72.221 & -72.211 & -72.242 & -72.222 & -72.244 & -72.223 & -72.245 & -72.224 \\
    0\% film & 5.83 & 19.34 & 6.18 & -68.302 & -68.309 & -68.315 & -68.315 & -68.315 & -68.315 & -68.315 & -68.315 \\
     2\% film & 5.94 & 18.58 & 6.03 & -68.198 & -68.205 & -68.042 & -68.043 & -68.213 & -68.213 & -68.213 & -68.214 \\
     \hline
    \end{tabular}
    \caption{
The in-plane ($l_a$) and out-of-plane ($l_c$) lattice constants, the distance $d$ between the Co layers, and the total energies per formula unit for Co$_{1/3}$NbS$_2$ bulk and the proposed thin film structure shown in Fig.\:\ref{fig:struct}. All structural parameters are obtained from the full relaxation using DFT for Co$_{1/3}$NbS$_2$ bulk and DFT+U for the thin film. For the spin states, we compare the collinear A-type AFM, FM, $1q-$AFM1, $1q-$AFM2, and the noncollinear $2q-$AFM1, $2q-$AFM2, $3q-$chiral \& $3q-$antichiral spin states. 
}
\label{tbl:energy}
\end{table*} 

\begin{figure}[!ht]
\vspace{-0.1cm}
\includegraphics[width=1.08\linewidth]{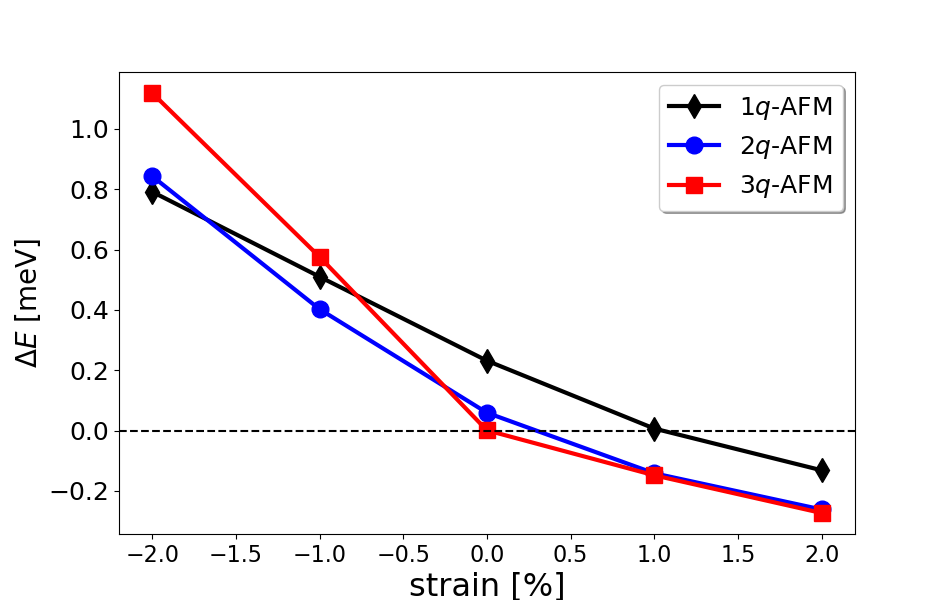}
\caption{The energy difference ($\Delta E$) of the magnetic states between the FM and AFM inter-layer magnetic couplings for $1q-$AFM (black diamond dots), $2q-$AFM (blue circle dots), and $3q-$AFM (red square dots). $1q-$AFM1(2), $2q-$AFM1(2), and $3q-$Chiral(Anti-chiral) states show the AFM(FM) inter-layer coupling between the Co layers. The anti-chiral $3q$ magnetic state is favored in the tensile (positively) strained thin film structure.
}
\label{fig:strain}
\end{figure}

First, we study the energetic stability of the various magnetic structures in the proposed Co bi-layer slab structure.
Our previous DFT calculation of Co$_{1/3}$NbS$_2$ showed that 
the lowest-energy spin configuration is the $3q$ non-coplanar AFM with the finite spin chirality $\chi$, though this energy is only $\sim$3meV (per formula unit) lower than the competing one such as the collinear $1q$ AFM (see Table\:\ref{tbl:energy}).
In this competing $1q$ AFM order, Co spins are aligned antiferromagnetically along one of the in-plane directions.
Another possible competing AFM order is $2q$ AFM, where coplanar noncollinear spins are obtained by removing the $z-$components of $3q$ AFM order. All of these collinear and noncollinear AFM orders are consistent with the location of the magnetic ordering peak measured in the early neutron scattering experiment on Co$_{1/3}$NbS$_2$~\cite{parkin_magnetic_1983}.
The A-type AFM order, where Co spins are aligned ferromagnetically in each layer but antiferromagnetically ordered along the $z-$direction, had higher energy ($\sim$24meV per formula unit) than the ground state one.
We find that a similar DFT calculation on the proposed (NbS$_2$)$_3$/Co/(NbS$_2$)$_3$/Co/(NbS$_2$)$_3$ thin film structure produces almost zero magnetic moments of Co ions due to the enhanced hybridization with the increased number of bands in NbS$_2$ layers near the Fermi energy. 
Inclusion of correlation effects is therefore necessary to reveal the possibly high-spin state of Co ions and the real-space topological effects due to the non-collinear spin texture.
We adopt the DFT+U method and obtain a high spin moment of 2.46$\mu_B$ per Co ion, which is similar to experimentally measured Co spin moment ($\sim2.73\mu_B$) in Co$_{1/3}$NbS$_2$~\cite{parkin_magnetic_1983}. 
Our DFT+U energy calculation 
shows that the $3q$ non-coplanar spin state has $\sim$13meV lower than the A-type AFM state, while its energy is almost degenerate to the competing $1q$ AFM state.  

Next, we study the dependence of the inter-layer magnetic coupling of the thin film  on strains using DFT+U. 
First, we relax the atomic positions of each strained structure using DFT+U while both in-plane and out-of-plane lattice constants are fixed.
One should note that the Co inter-layer distance $d$ is also allowed to relax since the out-of-plane lattice constant $l_c$ includes the vacuum layers.
Then, we compute the energies of A-type AFM, FM, $1q$ collinear AFM1(2), $2q$ coplanar \& noncollinear AFM1(2)  which ordered antiferromagnetically(ferromagnetically) along the $z-$direction, and $3q$ noncollinear chiral \& antichiral AFM structures for each crystal structure.
To study the inter-layer magnetic coupling, we compute the energy difference ($\Delta E$) between the ferromagnetically ordered structure ($1q-$AFM2, $2q-$AFM2, $3q-$AC) and the anti-ferromagnetically ordered one ($1q-$AFM1, $2q-$AFM1, $3q-$Chiral) along the $z-$direction.
Without any strain effects, $\Delta E$ is tiny for a11 $1q$, $2q$, and $3q$ cases, while $E_{FM}$ is almost $\sim$7meV lower than A-type $E_{AFM}$, even though the 3$q$ non-coplanar spin energy is still lower than the pure FM case by $\sim$6meV. We find that $\Delta E$ is systematically lower for the thin film case compared to the same spin states in Co$_{1/3}$NbS$_2$ bulk (see Table\:\ref{tbl:energy}). 
As the in-plane tensile strain is applied to the thin film structure, all strained structures ferromagnetically ordered along the $z-$direction become more favored, although the energy dependence on the strain can still be weak (see Fig.\:\ref{fig:strain}). 
This shows that strains can be an effective way to tune the interlayer magnetic coupling of layered materials such that the tensile strain can favor the ferromagnetically ordered spin configuration along the $z-$direction as the interlayer hopping between Co sites increases. 
%
Since the anti-chiral $3q$ magnetic structure is the ground state for the 2\% tensile strained film and it can possibly show the large magnetoelectric coupling, we use this 2\% tensile strained structure to compute the electronic structure (see details in Appendix A), the Berry curvature, and the orbital magnetization.





\section{Orbital magnetization}

\begin{figure}[!ht]
\vspace{-0.1cm}
\includegraphics[width=\linewidth]{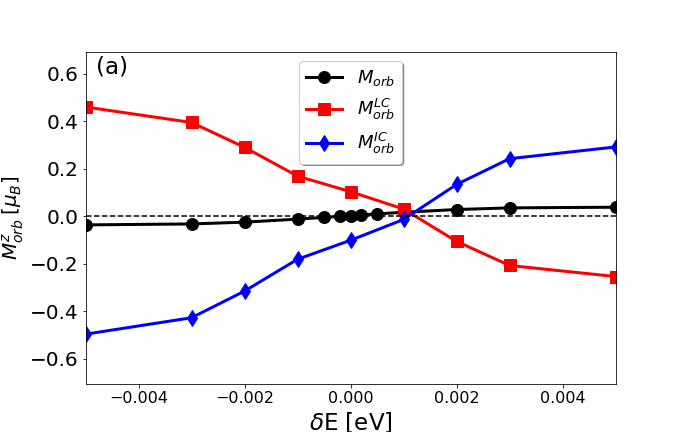}
\includegraphics[width=\linewidth]{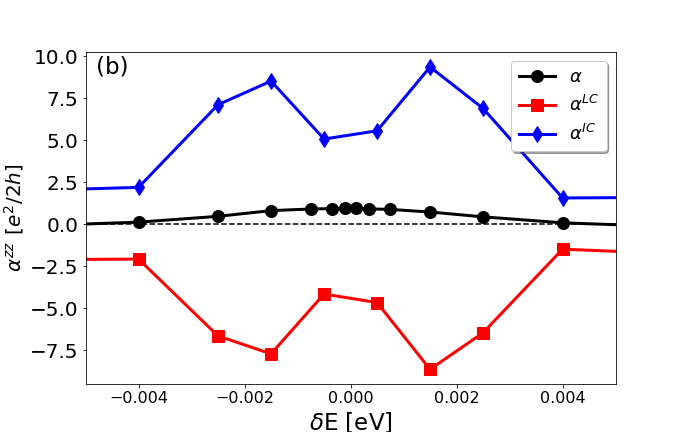}
\caption{(a) The orbital magnetization $M_{orb}$ per magnetic unit cell for the anti-chiral (AC) magnetic structure under the 2\% tensile strain calculated as a function of the orbital energy difference $\delta E$. The total $M_{orb}$ consists of the local circulation ($M^{LC}_{orb}$) and the itinerant circulation ($M^{IC}_{orb}$) terms. (b) The magnetoelectric coupling $\alpha$ computed from the linear slope of $M_{orb}$ vs $\delta E$. 
}
\label{fig:morb}
\end{figure}

\begin{figure}[!ht]
\vspace{-0.1cm}
\includegraphics[width=\linewidth]{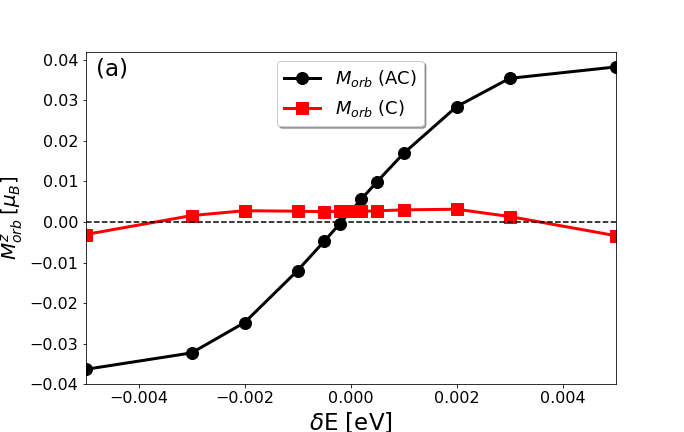}
\includegraphics[width=\linewidth]{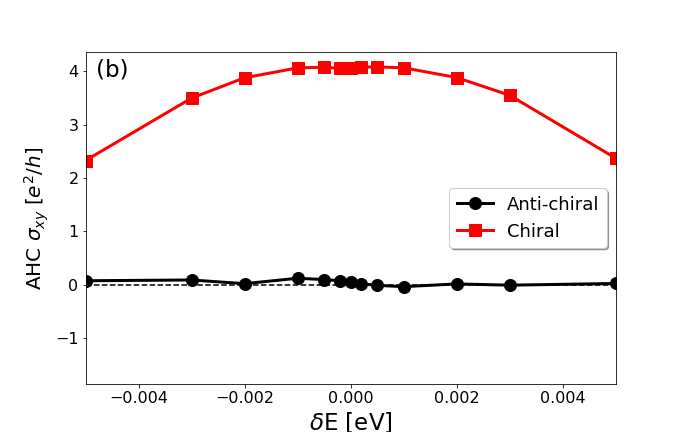}
\caption{(a) The orbital magnetic moment per  magnetic unit cell compared for the anti-chiral (AC) and the chiral (C) magnetic structures under the 2\% tensile strain. (b) The anomalous Hall conductivity $\sigma_{xy}$ computed using the Berry curvature for both structures. 
}
\label{fig:berry}
\end{figure}

We now compute the orbital magnetization $M_{orb}^z$ using the following formula obtained on the basis of the Wannier representation~\cite{PhysRevB.85.014435}:
\begin{eqnarray}
\label{eq:morb}
M_{orb}^z&=&\frac{e}{2\hbar}Im\sum_n\int\frac{d\mathbf{k}}{(2\pi)^3}f_{n\mathbf{k}}\cdot \nonumber\\
&&\langle\partial_{k_x} u_{n\mathbf{k}}|(\hat{H}_{\mathbf{k}}+\epsilon_{n\mathbf{k}}-2\epsilon_F)|\partial_{k_y} u_{n\mathbf{k}}\rangle,
\end{eqnarray}
where $\hat{H}_k$ is the (spin-polarized) Bloch Hamiltonian with eigenvalues $\epsilon_{n\mathbf{k}}$ and the cell-periodic Bloch functions $u_{n\mathbf{k}}$. $f_{n\mathbf{k}}$ is the Fermi function and $\epsilon_F$ is the Fermi energy. This Wannier-based implementation can accurately treat the orbital magnetization in the presence of the  Fermi surface,  as was shown in the cases of Fe, Co, and Ni ~\cite{PhysRevB.85.014435}. 
The total $M_{orb}$ consists of two  contributions: the ``local circulation" (LC) $M^{LC}_{orb} \sim\langle\partial_{k_x} u_{n\mathbf{k}}|(\hat{H}_{\mathbf{k}}-\epsilon_F)|\partial_{k_y} u_{n\mathbf{k}}\rangle$ and the ``itinerant circulation" (IC) $M^{IC}_{orb}\sim\langle\partial_{k_x} u_{n\mathbf{k}}|(\epsilon_{n\mathbf{k}}-\epsilon_F)|\partial_{k_y} u_{n\mathbf{k}}\rangle$, such that $M_{orb}=M^{LC}_{orb}+M^{IC}_{orb}$ ~\cite{PhysRevLett.95.137205}. 
The LC term originates from the bulk, while the IC term corresponds to the orbital magnetization associated with the surface current. 


Fig.\:\ref{fig:morb} (a) shows the orbital magnetization computed in the AC magnetic structure as a function of the electric field.
We assumed a constant $\mathbf{E}$ field which induces  linear change of the electric potential along the $z-$direction.
As a result, the change of the Hamiltonian due to the electric field is given by $\delta \hat{H}(\mathbf{E})=eE_z\cdot\hat{\mathbf{d}}$ where $\hat{\mathbf{d}}$ is the position operator along the $z-$axis. 
Both LC and IC terms are large in magnitude and depend linearly on the electric field. 
They have opposite signs and the total $M_{orb}$ has a slightly larger contribution from the IC term.
The AC magnetic structure has a small but finite $M_{orb}$ contribution even under zero field due to the lack of the inversion symmetry.
The change of $M_{orb}$ terms due to the electric field is linear near the zero-field limit as the on-site orbital energy shift near the Fermi energy induces the nearly linear band structure change, which  contributes to the Berry curvature and to the orbital magnetization.

The magnetoelectric coupling $\alpha^{zz}$ can be obtained from the slope of the orbital magnetization $M^z_{orb}$ (Fig.\:\ref{fig:morb}) with respect to $E_z$, i.e. $\alpha^{zz}=\partial M_{orb}^z/\partial E_z$.
Our estimate of the $\alpha$ term for Co$_2$(NbS$_2$)$_9$ in the AC magnetic structure produces a large value $\sim$0.9 in the units of $e^2/2h$. The $IC$ term for $\alpha^{zz}$ is almost $\sim$ 5.3 while the $LC$ term is $\sim$ -4.4, indicating that most of the $IC$ contribution is  compensated due to the $LC$ contribution in the total  magnetoelectric coupling.

For comparison, in the chiral magnetic structure, the dependence of $M_{orb}$ on the electric field is much weaker than in the AC magnetic case, as shown in Fig.\:\ref{fig:berry}a. On the other hand, the calculated Berry curvature in the chiral structure is much enhanced compared to AC, with some dependence of the electric field, indicating that the large AHE is originating from the topological band structure of this system.
The anomalous Hall conductivity of this chiral structure would be $\sim 1.4\:e^2/h$ per NbS$_2$ layer, a value similar to what we found previously from the first-principles calculation for bulk Co$_{1/3}$NbS$_2$ ~\cite{PhysRevMaterials.6.024201}.


\begin{figure}[!ht]
\vspace{-0.1cm}
\includegraphics[width=\linewidth]{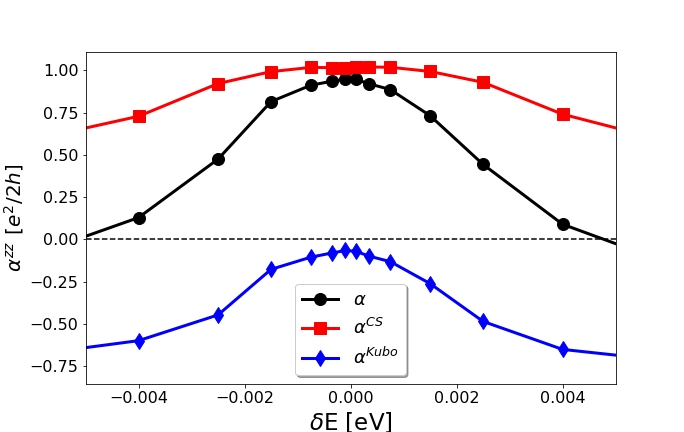}
\caption{The decomposition of the magnetoelectric coupling $\alpha^{zz}$ for the anti-chiral magnetic structure into the Chern-Simon term ($\alpha^{CS}$) and the Kubo-like term ($\alpha^{Kubo}$).}
\label{fig:alpha}
\end{figure}

It is also possible to derive analytically the formula for $\alpha$ directly from Eq.\:\ref{eq:morb}. Resulting formula  contains two contributions: the Chern-Simons magnetoelectric coupling ($\alpha^{CS}$) obtained from the $M_{orb}$ term depending on the electric field explicitly and the ``Kubo-like" term ($\alpha^{Kubo}$) originating from the first-order changes of the non-interacting Hamiltonian or wavefunctions, which depends on the electric field implicitly~\cite{Malashevich_2010}. 
The $\alpha^{CS}$ originates from the itinerant circulation term, $\mathbf{M}^{IC}(\mathbf{E})$, which depends explicitly on the $\mathbf{E}$ field:
\begin{eqnarray}
\left(\frac{dM_{orb}}{dE_z}\right)_{CS}&=&\frac{e}{2\hbar}\frac{1}{(2\pi)^3}Im\sum_n\int d^3kf_{n\mathbf{k}}\langle\partial_{x} u_{n\mathbf{k}}|e\hat{z}|\partial_{y} u_{n\mathbf{k}}\rangle \nonumber \\ 
&\simeq& \frac{e^2}{2h}\frac{d}{l_c}\frac{1}{2\pi}\sum_n\int d^2kf_{n\mathbf{k}} \left[\Omega_n^{T}(\mathbf{k})-\Omega_n^{B}(\mathbf{k})\right]
\label{eq:alpha_CS}
\end{eqnarray}
where $\hat{z}$ is the position operator along the $z-$direction, $d$ is the $z-$position of the Co layer, $l_z$ is the lattice constant along the $z-$direction, and $\Omega_n^{T(B)}(\mathbf{k})$ is the Berry curvature of the band $n$ with the momentum $\mathbf{k}$ contributed by the top (bottom) surfaces of the system.  
The integrated Berry curvature is directly related to the anomalous Hall conductivity.
We see that the Berry curvature contributions from the top and bottom layers enter with opposite signs.
A similar expression was also derived to obtain the magnetoelectric coupling for other layered systems with layer-dependent Chern numbers~\cite{Qiu2025,Lu2025}.
The center of the top (bottom) component is roughly located at the top (bottom) Co layer, i.e., $\langle z \rangle \sim \pm d_{Co}$ and the magnitude is $\sim l_z/4$, where $l_z$ is the lattice constant along the $z-$direction.
Since the total Berry curvature for the anti-chiral magnetic structure is almost zero regardless of the $\mathbf{E}$ field (see Fig.\:\ref{fig:berry}b), we argue that the Berry curvature difference between top and bottom components is estimated to equal the total Berry curvature of the chiral structure, i.e., $(\Omega(\mathbf{k})^T-\Omega(\mathbf{k})^B)_{AC}\sim \Omega(\mathbf{k})_{chiral}$.

Fig.\:\ref{fig:alpha} shows the decomposition of the $\alpha^{zz}$ into the $\alpha^{CS}$ term obtained using Eq.\:\ref{eq:alpha_CS} and the remaining $\alpha^{Kubo}$ contribution.
Under the zero field, most of the $\alpha^{zz}$ contribution is attributed to the $\alpha^{CS}$ term which is directly related to the layer-dependent Berry curvature and the surface anomalous Hall conductivity of the chiral magnetic structure. This suggests that a large axion coupling can be naturally realized in various bi-layer materials with the large anomalous Hall conductivity if by tuning  the inter-layer magnetic coupling they can be switched from chiral to antichiral magnetic state. 

\section{Conclusions}
In summary, based on first-principles calculations, we propose that a very large magneto-electric coupling, $\alpha^{zz}$, can be realized in the Co-intercalated 1-H NbS$_2$ thin film structure.
The Co ions form a triangular lattice with a non-coplanar 3$q$ spin structure exhibiting a finite scalar spin chirality. The Co bi-layers in the proposed thin film system can have the same or opposite signs of the spin chirality, depending on the sign of the inter-layer spin exchange interaction. The latter can be tuned by application of the in-plane strain.
While the same-sign chiral structure produces large net Berry curvature leading to the large anomalous Hall conductivity similarly to the bulk Co$_{1/3}$NbS$_2$ case, the anti-chiral structure produces the large topological magneto-electric effect originating from the layer-dependent Berry curvature even though the total Berry curvature of the system is negligible.
Our first-principles calculation predicts that the magneto-electric coupling $\alpha^{zz}$ can be as large as 0.9 $e^2/2h$.
We expect this large magneto-electric coupling can also be obtained for other even numbers of Co layers beyond the bi-layers considered here.
Our result opens a new avenue to control  intercalated transition-metal dichalcogenides, revealing novel topological physics, including the anomalous Hall effect and the magneto-electric (axion) states, tunable by strains.

\section*{Acknowledgement}
This work was supported by the Materials Science and Engineering Division, Basic Energy Sciences, Office of Science, US Department of Energy. We gratefully acknowledge the computing resources provided on Bebop and Improv, a high-performance computing cluster operated by the Laboratory Computing Resource Center at Argonne National Laboratory.


\newpage

\begin{figure*}[!ht]
\vspace{-0.1cm}
\includegraphics[width=0.35\linewidth]{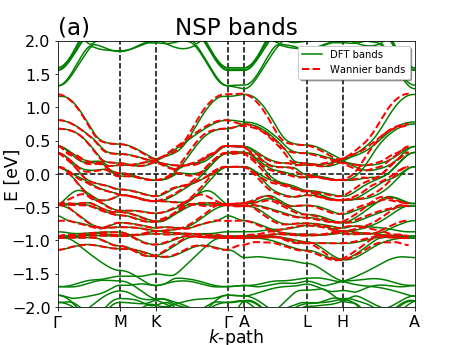}
\hspace{-0.73cm}
\includegraphics[width=0.35\linewidth]{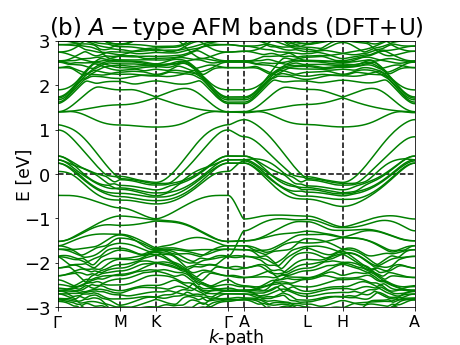}
\hspace{-0.73cm}
\includegraphics[width=0.35\linewidth]{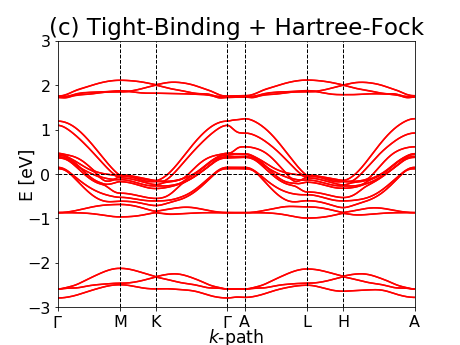}
\caption{(a) The non-spin-polarized (NSP) DFT band structure of Co$_2$(NbS$_2$)$_9$ (The dashed line is the Wannier interpolated band), (b) The A-type anti-ferromagnetic (AFM) DFT+U band structure of Co$_2$(NbS$_2$)$_9$, (c) The AFM bands fitted from the NSP tight-binding model with spin exchange potentials.
}
\label{fig:band}
\end{figure*}


\section*{Appendix A: Electronic structure}

To simulate the spin-polarized bands and perform Berry curvature calculations, we adopted the Wannier-interpolated band technique based on the tight-binding Hamiltonian to increase the number of $k-$points in the simulation.
First, we construct the tight-binding Hamiltonian from the non-spin-polarized (NSP) band structure of Co$_2$(NbS$_2$)$_9$ obtained from DFT~\cite{vasp1,vasp2} using Co 3$d$ and Nb 4$d_{z^2}$ maximally localized Wannier orbitals~\cite{Wannier90_2012}.
Our Wannier-interpolated bands based on the tight-binding Hamiltonian reproduce the DFT bands almost exactly within the energy window of [-1.5, 1.5] eV, as shown in Fig.\:\ref{fig:band}a.
To include the slab geometry to Co$_2$(NbS$_2$)$_9$, we ignore the inter-cell hoppings along the $z-$direction from the tight-binding Hamiltonian.

Then, we analyze the A-type AFM band structure of Co$_2$(NbS$_2$)$_9$ obtained from DFT+U, as shown in Fig.3 middle.
While the itinerant bands near the Fermi energy are mostly Nb 4$d$ characters, Co $3d$ bands are split by the Hubbard $U$ interaction within DFT+U.
Since each Co ion is surrounded by six sulfur ions following the trigonal symmetry, the crystal field of the Co $3d$ orbital is split into two doublets and one singlet under local $D_{3h}$ symmetry.
The doublet of the $d_{xz}$/$d_{yz}$ orbitals have higher energy than the $d_{x^2-y^2}$/$d_{xy}$ orbitals since their orbital lobes point toward sulfur ions.
As a result, the $d_{xz}$/$d_{yz}$ and $d_{z^2}$ orbitals are almost half-filled and strongly split due to the Hubbard interaction ($U$=5eV) resulting in the relatively flat bands located $\sim$1-2eV above the Fermi energy and near -3eV below the Fermi energy. 
The $d_{x^2-y^2}$/$d_{xy}$ orbitals are hybridized weakly with the neighboring sulfur ions and almost fully occupied with bands located near -1eV below the Fermi energy.
Nb bands are less affected by the Co Hubbard $U$ and they exhibit dispersive feature near the Fermi energy forming the metallic Fermi surface.

To incorporate the spin-dependent potential effect to the NSP tight-binding bands, we adopt the similar Hartree-Fock approximation as we used in our previous study of Co$_{1/3}$NbS$_2$.
The following on-site spin-dependent Hartree-Fock Hamiltonian ($\hat{H}_U$) is added to the local Co site of the NSP tight-binding Hamiltonian:
\begin{equation}
\hat{H}_U \simeq \sum_{i\alpha}\sum_{\sigma\sigma'} \left(V_{\alpha}\cdot\delta_{\sigma\sigma'}+\mathbf{J}_{\alpha}\cdot\mathbf{\sigma}_{\sigma\sigma'}\right)\hat{c}^{\dagger}_{i\alpha\sigma}\hat{c}_{i \alpha\sigma'},
\label{eq:HF}
\end{equation}
where $V_{\alpha}\simeq \sum_{\beta}\bar{U}_H\langle\hat{n}^i_{\beta}\rangle$ and $\mathbf{J}_{\alpha}\simeq \sum_{\beta}\bar{U}_F \langle\hat{\mathbf{S}}^i_{\beta}\rangle$ are the effective spin-dependent potential for the orbital $\alpha$. This orbital-dependent Hubbard interaction terms are similarly adopted within the DFT+U calculation. For our thin film structure, we used $V_{\alpha}\sim -0.1$eV for $d_{xz}$/$d_{yz}$ and $d_{z^2}$ orbitals and $V_{\alpha}\sim -2.5$eV for $d_{x^2-y^2}$/$d_{xy}$ orbitals. 
The magnitude of the $\mathbf{J}_{\alpha}$ term is set to 2.1eV for all Co d orbitals to ensure the strong spin split bands obtained in DFT+U.
Our tight-binding band structure combined with the Hartree-Fock Hamiltonian in Eq.\:\ref{eq:HF} indeed captures essential features of the most itinerant Nb bands near the Fermi energy as well as the localized and flat Co 3$d$ band positions when compared to the DFT+U band structure.

\section*{Appendix B: Schematic spin configurations}

\begin{figure*}[!ht]
\vspace{-0.1cm}
\includegraphics[width=0.7\linewidth]{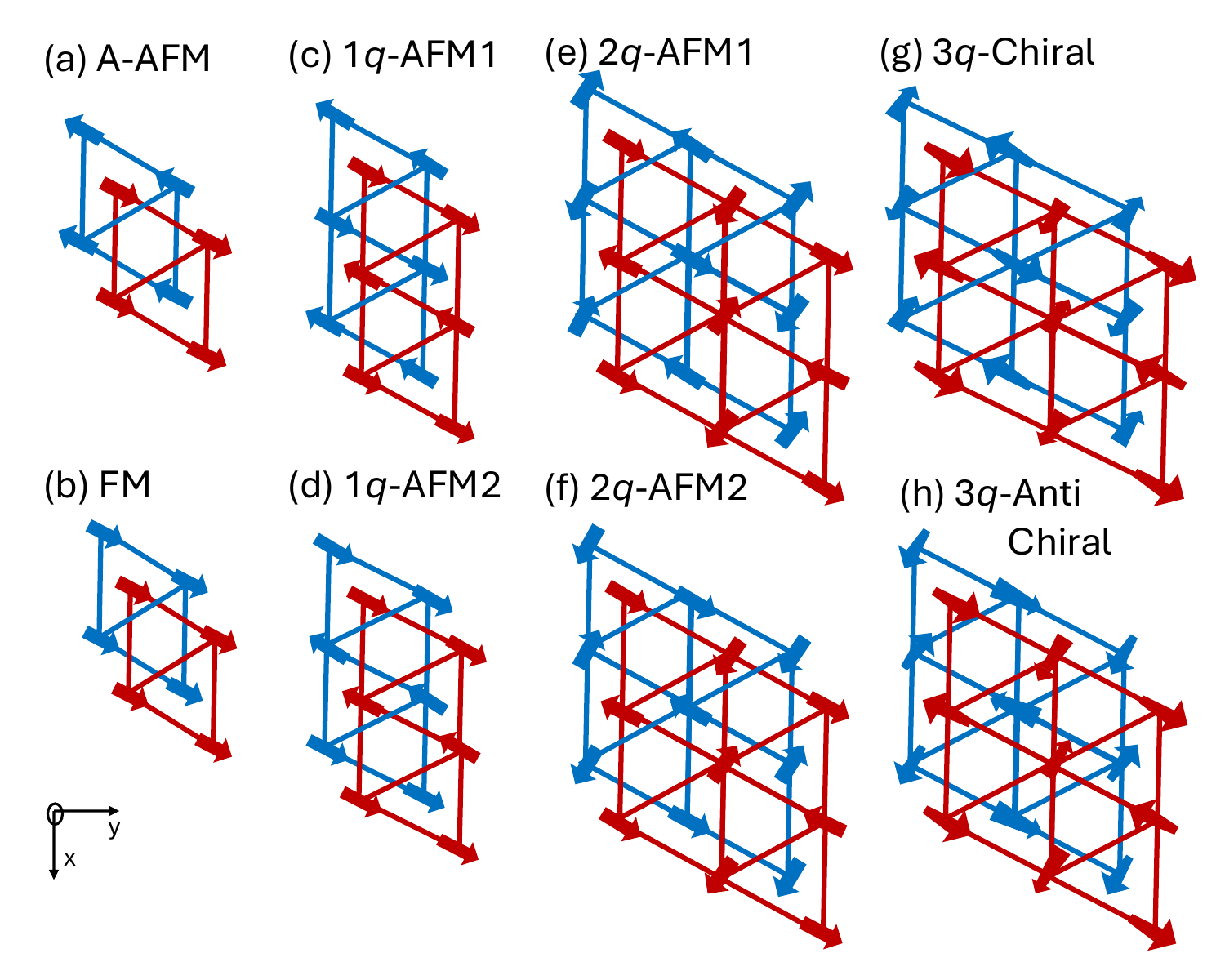}
\caption{The schematic spin textures for possible spin configurations discussed in the Table I of the main text.
}
\label{fig:spin_fig}
\end{figure*}

We show the schematic spin textures for possible spin configurations in Fig.\:\ref{fig:spin_fig}, which were also discussed in Table I of the main text.

\section*{Appendix C: Energy barrier between $3q-$Chiral and $3q-$AntiChiral magnetic states}

\begin{figure*}[!ht]
\vspace{-0.1cm}
\includegraphics[width=0.7\linewidth]{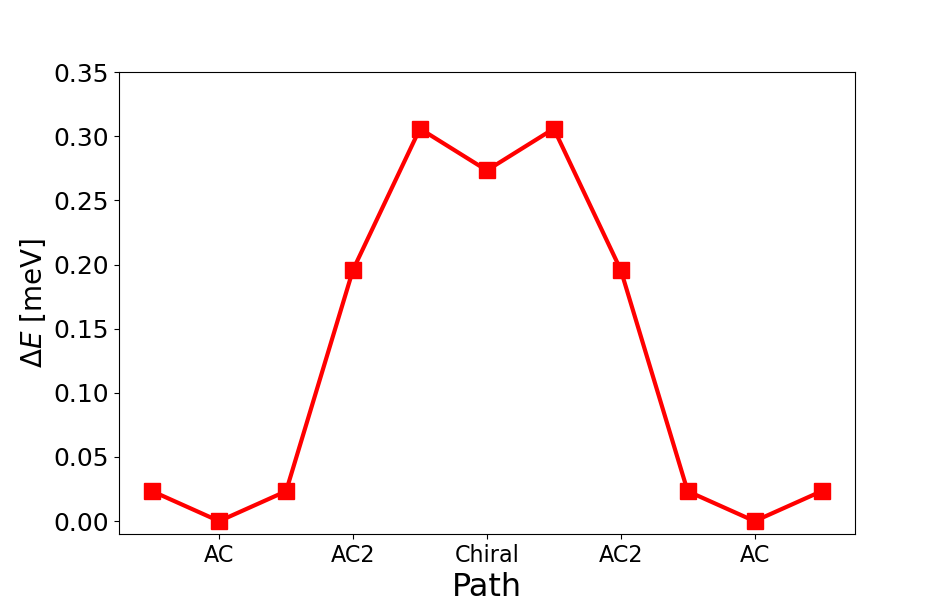}
\caption{The energy of $3q$ noncollinear magnetic states along the path switching between $3q-$Chiral and $3q-$AntiChiral states for the 2\% tensile strained Co$_2$(NbS$_2$)$_9$.
}
\label{fig:Epath}
\end{figure*}

While our main results show that the $3q-$AntiChiral (AC) magnetic state will be the ground state for the 2\% tensile strained Co$_2$(NbS$_2$)$_9$ thin film, the $3q-$Chiral state still needs to be metastable for the switching mechanism between two magnetic states. Here, we plot the energy of the $3q-$magnetic states along the path connecting the $3q-$AC and the $3q-$Chiral states (Fig.\:\ref{fig:Epath}). To switch from the $3q-$AC to the $3q-$Chiral state, one first needs to rotate the spins by 180$^{\degree}$ about the $z-$axis ($C^z_2$) in one of the Co layers and obtain the $3q-$AC2 state. Then, one needs to apply the mirror symmetry to the spins in this Co layer about the $x-y$ plane ($M^z$) to change the sign of the spin chirality. Our result shows that a small energy barrier is present and the $3q-$Chiral state is meta-stable, while the spin chirality is changed under the application of the mirror symmetry. Although this energy barrier is still very small, one can possibly use the strain to tune this energy barrier and the magnetic ground state. We also note that this discussion pertains only to the low-temperature switching between two phases. When cooling from high temperatures, there will be no barrier for the system to enter the chiral phase.

\bibliography{main}

\end{document}